# Structural and Magnetic Properties of Molecular Beam Epitaxy Grown Chromium Selenide Thin Films

Anupam Roy [1]*, Rik Dey [1], Tanmoy Pramanik [1], Amritesh Rai [1], Ryan Schalip [1], Sarmita Majumder [1], Samaresh Guchhait [2], and Sanjay K Banerjee [1]

[1] Microelectronics Research Center, The University of Texas at Austin, Austin, Texas 78758, USA

[2] Department of Physics and Astronomy, Howard University, Washington, DC 20059, USA

*Address correspondence to: anupam@austin.utexas.edu.

## ABSTRACT

Chromium selenide thin films were grown epitaxially on $Al_2O_3$(0001) and Si(111)-(7×7) substrates using molecular beam epitaxy (MBE). Sharp streaks in reflection high-energy electron diffraction and triangular structures in scanning tunneling microscopy indicate a flat smooth film growth along the *c*-axis, and is very similar to that from a hexagonal surface. X-ray diffraction pattern confirms the growth along the *c*-axis with *c*-axis lattice constant of 17.39 Å. The grown film is semiconducting, having a small band gap of about 0.034 eV, as calculated from the temperature dependent resistivity. Antiferromagnetic nature of the film with a Néel temperature of about 40 K is estimated from the magnetic exchange bias measurements. A larger out-of-plane exchange bias, along with a smaller in-plane exchange bias is observed below 40 K. Exchange bias training effects are analyzed based on different models and are observed to be following a modified power-law decay behavior.

## I. Introduction

Binary chromium-based chalcogenides exhibit various interesting physical properties with a wide variation in electrical and magnetic properties. A small change in composition changes the physical properties and makes them more fascinating as a material system to study. Wontcheu *et al.* have shown the effect of anion substitution on the structural and magnetic properties of chromium chalcogenides [1]. The chromium-selenium system is a large family of compounds with large varieties of stable stoichiometries [*e.g.*, $Cr_{1-x}Se$, $Cr_2Se_3$, $Cr_3Se_4$, $Cr_5Se_8$, $Cr_7Se_8$, *etc*.]. All of these compounds have NiAs-type crystal structure. Due to incomplete *d*-orbitals of the transition metal, these NiAs-type structures show interesting magnetic and electrical properties [2]. Different compounds of chromium selenides differ on the Cr-vacancies that occur in every second metal layer. Thus, every alternate layer of metal-deficient and metal-rich layers stack along the *c*-axis [3-7]. Magnetic properties of bulk $Cr_2Se_3$ have been studied extensively before, and it has been shown to be an antiferromagnet below the Neel temperature, $T_N$ ~ 43 K. Also, an order-order transition occurs at ~ 38 K between the low-temperature and high-temperature antiferromagnetic $Cr_2Se_3$ structures, as observed from neutron diffraction studies [6-9]. Because of Cr vacancies in alternate layers, the moment associated with Cr atoms located on two different layers are different due to different neighboring environment and this leads to the complexity in the magnetic structure below $T_N$.

Previously, chromium selenide systems have been studied to investigate their suitability as thermoelectric material for intermediate-temperature applications [10-14], intermediate temperature power generation [13], electrochemical sensors [15], *etc*. Several groups have studied the structural, magnetic, electrical and thermoelectric properties of single crystal $Cr_{2+x}Se_{3-x}$ compounds grown using solid state reaction method [10-13,16], soft chemical and hydrothermal synthesis [15,17,18] and chemical vapor transport method [4,14,19-21]. However, the studies focus mostly on the improvement in thermoelectric properties of transition-metal-doped bulk samples of $Cr_2Se_3$. The epitaxial growth and different physical properties of $Cr_2Se_3$ thin films are yet to be explored in detail. Molecular beam epitaxy (MBE) is a highly specialized technique used to grow ultra-high purity large-area epitaxial thin films with abrupt interfaces and with precise control over their thicknesses. Compared to other growth techniques, MBE offers greater control to incorporate dopants in thin films. This makes it even more suitable growth method, as the electrical, magnetic and thermoelectric properties of this material system can be largely varied with addition of transition-metal/chalcogen dopants [1, 9-13, 21,22].

In this work, we report the epitaxial growth of $Cr_2Se_3$ thin films under ultra-high vacuum (UHV) directly on $Al_2O_3(0001)$ and Si(111)-(7×7) surfaces using MBE. Interestingly, we show that the growth occurs along (001) direction (*c*-axis). We present the details of growth, structural, electrical and magnetic properties characterized by several *in situ* and *ex situ* techniques, *e.g.*, reflection high energy electron diffraction (RHEED), x-ray diffraction (XRD), scanning tunneling microscopy (STM), x-ray photoelectron spectroscopy (XPS), magneto-transport measurements, *etc*. We report the exchange bias training behavior of the epitaxial $Cr_2Se_3$ thin film coupled with a ferromagnet, characterized using magnetoresistance (MR) measurements, and analyzed in detail using different models.

## II. Experimental Method

**Growth**: $Cr_2Se_3$ films were grown in a custom-built MBE growth system (Omicron, Germany) under ultra-high vacuum (UHV) conditions (base pressure $\sim 1\times10^{-10}$ mbar). Details of the system has been described elsewhere [23]. A reflection high-energy electron diffraction (RHEED) setup is attached to the MBE system for *in situ* monitoring of surface reconstruction and growth. Substrates used in the experiment were insulating *c*-axis $Al_2O_3(0001)$ and P-doped n-type Si(111) wafers (oriented within ±0.5°) with a resistivity of 1-20 Ω-cm. After the substrates were precleaned in acetone and isopropanol, the substrates were introduced into the UHV chamber. Atomically clean, reconstructed Si(111)-(7×7) surfaces were prepared by the usual heating and flashing procedure [24]. Single crystal *c*-$Al_2O_3(0001)$ substrates were prepared by resistive heating at 600 °C for 3 hours followed by 700 °C for 30 min. Clean substrate surfaces were examined by *in situ* RHEED. Chromium and selenium fluxes generated by *e*-beam evaporator and effusion cell, respectively, were co-deposited on the substrates at an elevated substrate temperature of about 340 °C. The chamber pressure during growth never exceeded $1\times10^{-9}$ mbar and the $Se_2$/Cr BEP (beam equivalent pressure) flux ratio was kept at about 15. Several samples with thicknesses varying from 5 nm to 25 nm were grown and typical growth rate of $Cr_2Se_3$ films was about 0.1 nm/min.

**Characterization**: Post-growth investigations of the samples were carried out by *in situ* RHEED operated at 13 kV, scanning tunneling microscopy (STM) at room temperature (RT) in the constant current mode, and X-ray photoelectron spectroscopy (XPS) with monochromatic Al-K$\alpha$ source ($h\nu$ = 1486.7 eV) operating at 15 kV. A Philips X-Pert X-ray diffraction (XRD) system equipped with a Cu X-ray filament source and a PW-3011/20 proportional detector was used for the *ex situ* XRD measurements.

**Electrical and Magnetic Measurements**: Transport measurements were carried out with 9 T Quantum Design physical property measurement system (PPMS) combined with vibrating sample magnetometry (VSM) capable of cooling samples down to ~ 2 K. The measurements were conducted using standard Van der Pauw method with indium dot contacts at the four corners of the large area rectangular samples.

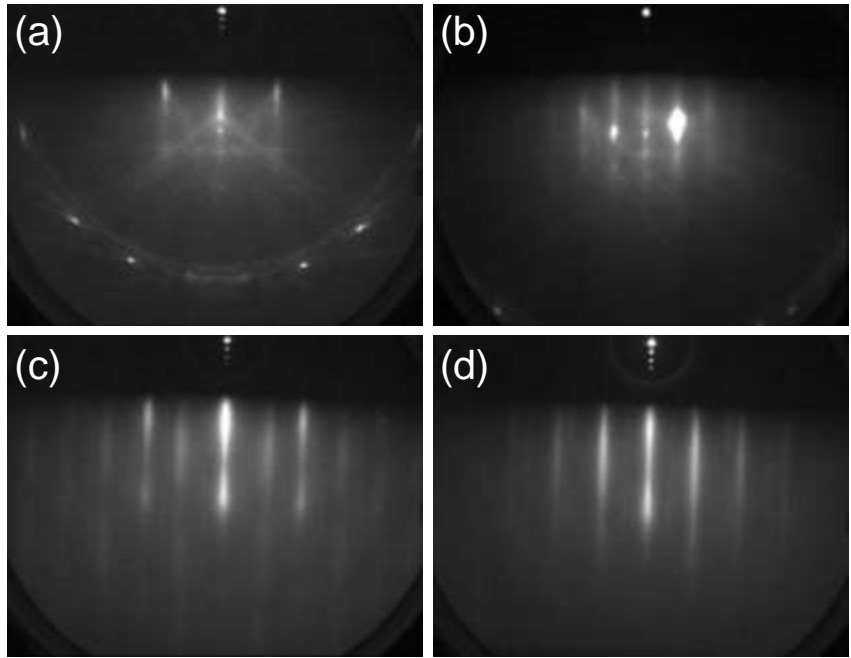

**Figure 1:** RHEED images following $Cr_2Se_3$ growth on $Al_2O_3(0001)$ surfaces. (a) & (b) RHEED patterns from a clean $Al_2O_3(0001)$ surface with the incident electron beam along [1 0 -1 0] and [1 1 -2 0] orientations of $Al_2O_3$, respectively. (c) & (d) Corresponding RHEED patterns from the same surface following 15 nm of $Cr_2Se_3$ growth.

## III. Results and discussions

### A. Growth and Characterizations

Bulk crystal growth of $Cr_2Se_3$ has been achieved previously by various methods, *e.g.*, ceramic method [6-9], solid state reaction method [10-13,16], soft chemical and hydrothermal synthesis [15,17,18] and chemical vapor transport method [4,14,19-21]. Using MBE, here we have studied the growth of $Cr_2Se_3$ thin films of different thicknesses directly on UHV-cleaned $Al_2O_3(0001)$ and $Si(111)$-$(7\times7)$ substrates without any buffer layer.

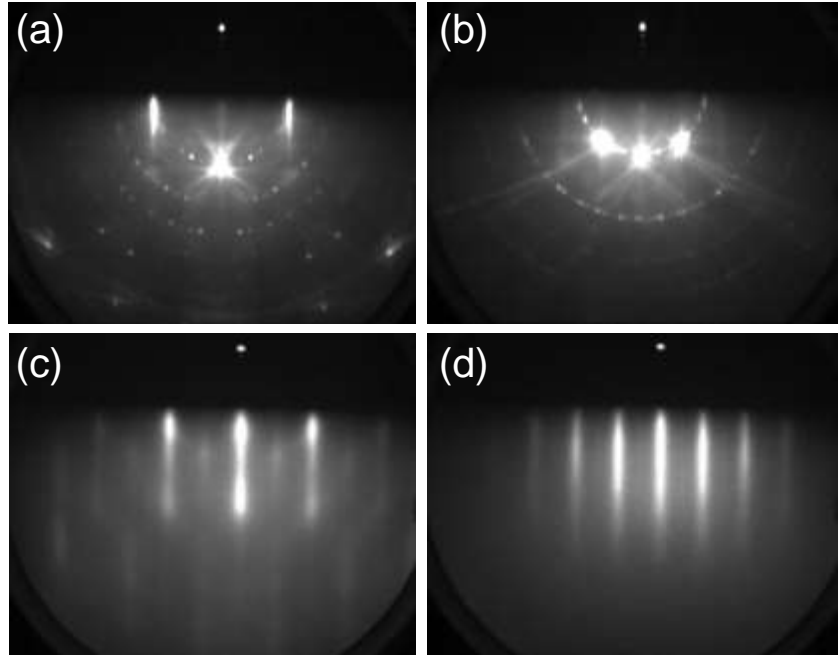

**Figure 2:** RHEED images following $Cr_2Se_3$ growth on $Si(111)$-$(7\times7)$ surfaces. (a) & (b) RHEED patterns from a clean $(7\times7)$ surface reconstruction from $Si(111)$ substrate along [1 1 -2] and [1 -1 0] orientations of Si, respectively. (c) & (d) Corresponding RHEED patterns from the same surface following 25 nm of $Cr_2Se_3$ growth.

Figures 1 & 2 display the RHEED patterns observed for the epitaxial $Cr_2Se_3$ thin film on $Al_2O_3(0001)$ and $Si(111)$-$(7\times7)$ substrates. Insulating crystalline sapphire substrates were chosen due to their hexagonal surface symmetry and to enable electrical measurements of the as-grown films. Figure 1 shows the RHEED images of the substrate before and after growth. Figures 1(a) & (b) show the RHEED patterns from the clean $Al_2O_3(0001)$ substrate along $[1\ 0\ -1\ 0]_{Al2O3}$ and $[1\ 1\ -2\ 0]_{Al2O3}$ directions, respectively, after a preheat treatment. Kikuchi lines are clearly observed in the RHEED patterns in Figs. 1(a) & 1(b) indicating a clean and smooth morphology of the substrate. Corresponding RHEED patterns observed after the growth are shown in Figs. 1(c) & 1(d). RHEED patterns from the sapphire substrate disappear completely within a few minutes of the growth at elevated temperature and resulted in sharp streaky RHEED patterns. In addition, half-order reconstructions are observed in the RHEED patterns reflecting a high crystal quality of the grown film. Streaky RHEED patterns are maintained throughout the growth process following the substrate surface crystal symmetry and the RHEED features are sensitive to both sample and beam orientation.

Similar growth has been achieved on Si(111) substrates, as shown in Fig. 2. RHEED patterns from reconstructed $Si(111)$-$(7\times7)$ surface are shown in Fig. 2(a) for the electron beam along $[1\ 1\ -2]_{Si}$ direction and in Fig. 2(b) for $[1\ -1\ 0]_{Si}$ incidence. Corresponding RHEED patterns, following epitaxial growth of

Cr$_2$Se$_3$ films, are shown in Figs. 2(c) & 2(d). Sharp streaky patterns in RHEED, for the growth on both Al$_2$O$_3$(0001) and Si(111)-(7×7) substrates, suggest well-structured film growth with high crystalline quality and smooth surface morphologies. Several samples with different thicknesses prepared on both the substrates show similar RHEED patterns, and for all of them, the RHEED patterns were maintained throughout the entire growth process (as also shown in Fig. S1 in the Supplemental Material [25] which includes Ref. [26-48]). Similar to the growth on Al$_2$O$_3$(0001) substrates, half-order reconstruction lines are present in the RHEED patterns [Figs. 2(c) & 2(d)]. Furthermore, the growth occurs along the *c*-axis (along (001) direction), as expected for the growth of a hexagonal thin film on an hcp(0001) or fcc(111) substrate. Similar (001)-oriented hexagonal thin film growth on fcc(111) surfaces also has been reported for Cr$_2$Te$_3$(0001), Bi$_2$Te$_3$(0001) and Bi$_2$Se$_3$(0001) on Si(111) substrates [23,49,50].

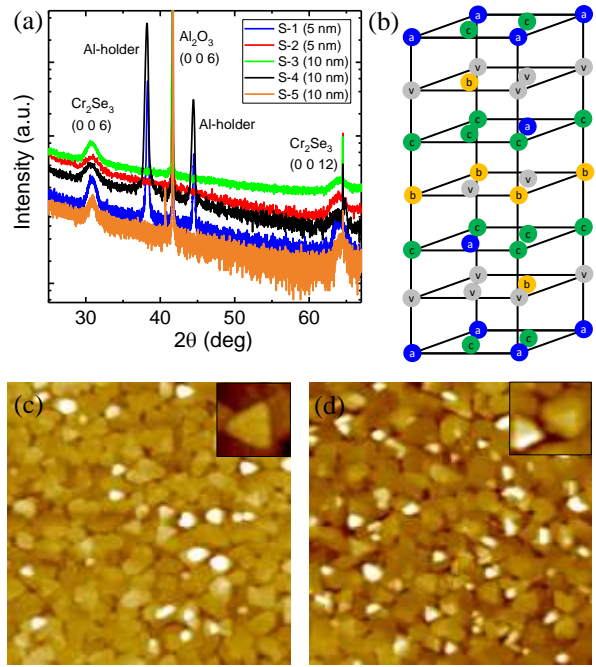

**Figure 3:** (a) XRD patterns from single crystal Cr$_2$Se$_3$ thin film of different thicknesses. The pattern shows that the growth is along (001) direction. (b) Rhombohedral crystal structure for Cr$_2$Se$_3$ where Se atoms are abbreviated for simplicity. Cr atoms are shown at three different Wyckoff sites ('a', 'b', 'c'), and vacancy sites are shown as 'v' (grey). (c) & (d) *In situ* STM studies of 5 nm epitaxial Cr$_2$Se$_3$ thin films grown on Si(111)-(7×7) surfaces. Several triangular features observed are reflecting the influence of substrate crystal symmetry. (Scan area: 300 × 300 nm$^2$, bias voltage: -2.5 V, tunneling current: 0.7 nA). Insets (25 × 25 nm$^2$) show (c) one such triangular domain and (d) triangular domains stack up as multiple layers.

XRD is used to evaluate the structure of the films and to confirm their epitaxial nature. Figure 3(a) shows the XRD patterns from Cr$_2$Se$_3$ thin films of different thicknesses grown on Al$_2$O$_3$(0001) substrates and shows characteristic peaks that correspond to diffraction from (00*l*) family of planes of the Cr$_2$Se$_3$ film. The sharp peak at $2\theta = 41.7°$ corresponds to reflection from the (006) plane of the Al$_2$O$_3$(0001) substrate. The XRD pattern agrees very well with the NiAs-type crystal of Cr$_2$Se$_3$ with hexagonal structure [ICSD

Collection Code 42705, space group R-3 (148)]. The crystal structure of rhombohedral $Cr_2Se_3$ is shown in Fig. 3(b). XRD peaks corresponding to the planes (0 0 6) and (0 0 12) of $Cr_2Se_3$ are indexed in Fig. 3(a). The absence of peaks other than the (0 0 *l*) family confirms the epitaxial growth along the *c*-axis of the sapphire substrates. XRD pattern also rules out any significant presence of any impurities and other known phases of chromium selenide. The extracted *c*-axis lattice constant of $Cr_2Se_3$ film is 17.39 Å, almost invariable for different samples (within 0.2%) and matches closely with the bulk crystal value of 17.38 Å [4].

Figures 3(c) & 3(d) show *in situ* STM studies of the surface of $Cr_2Se_3$ thin films grown on Si(111)-(7×7) surfaces. The structures are characteristically triangular shaped, reflecting the hexagonal crystal structure along the (001) direction. Because of the three-fold crystal symmetry of Si(111) substrate, formation of equilateral triangles is natural. Both hcp(0001) and fcc(111) surfaces have similar hexagonal Bravais lattice and they differ only in the registry of the third layer [51]. Compared to previous studies of bulk crystals grown *via* solid state reaction method [11,12], we notice diminished sizes of the structures in a thin film (~ 20-40 nm). Insets show the close-up shape of triangular structures. Closer examination of the triangular domain reveals that the shapes are truncated triangular or triangular hexagon. For hexagonal crystal structures, this happens when there is a mismatch in the rate of advancement of adatoms between the two edges during growth [52,53]. Inset of Fig. 3(d) illustrates multiple triangular domains stack up on top of each other. Similar observation of truncated triangular structures has also been made in case of hexagonal $Cr_2Te_3$ thin films grown epitaxially on Si(111)-(7×7) surfaces [49].

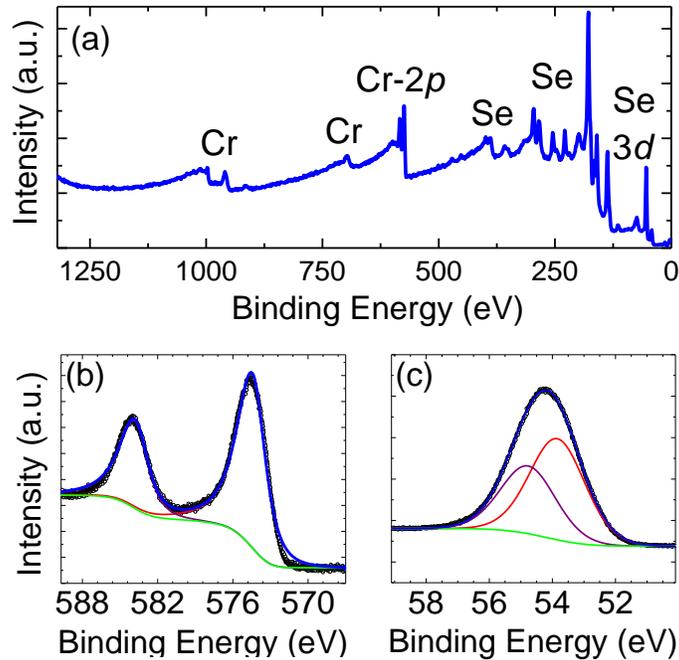

**Figure 4:** *In situ* high-resolution XPS spectra of $Cr_2Se_3$ thin films. (a) Survey scan from $Cr_2Se_3$ film. (b) and (c) are Cr-2*p* and Se-3*d* core-level photoemission spectra from 25 nm of $Cr_2Se_3$ thin film, respectively, show chemical shifts corresponding to homogeneous phases of hexagonal $Cr_2Se_3$. A Se/Cr ratio of about 1.5 is extracted from the area fit (solid lines) to the experimental data (○).

The elemental compositions of the grown films were examined by *in situ* XPS. Figure 4 (a) shows the XPS survey scans of the $Cr_2Se_3$ thin films. All the major peaks have been identified and assigned to Cr and Se. XPS also confirms the film to be free from the presence of any other elements as impurities. High-resolution XPS analysis of the sample finds peaks at Cr-2p and Se-3d edges, as shown in Figs. 4(b) and (c), respectively. The Cr-2p spectrum was fitted using an asymmetric peak shape due to $Cr_2Se_3$ being a narrow-bandgap semiconductor. The CasaXPS function LA(1.4,2,2), taken from a previous study on conductive Zn and its oxides, provided an excellent match [54,55]. The binding energies corresponding to the $Cr-2p_{3/2}$ and $Cr-2p_{1/2}$ peaks are at 574.4 eV and 583.8 eV, respectively, giving a doublet splitting of 9.4 eV. Binding energy corresponding to Cr-3s peak at 74.3 eV also matches very well. The positions of the Cr peaks are found to be in good agreement with previous literature values for $Cr_2Se_3$ [56]. Similarly, the Se-3d spectrum is also well fitted using GL(30) peak shapes and its peak location (54.2 eV) matches very well with previous reports [56]. Using the integrated areas under the peaks, a Se/Cr ratio of about 1.5 is obtained.

### B. Electrical properties

*3d* transition metal chalcogenides show various intriguing electrical and magnetic properties. Different combination of transition metal and chalcogen atoms can lead to various distinct and complex electrical and magnetic characteristics. For example, $Cr_{1-\delta}Se$ are mostly antiferromagnetic and semiconducting for $0 \leq \delta \leq 0.33$, whereas $Cr_{1-\delta}Te$ are all ferromagnetic with metallic conductivities for $0 \leq \delta \leq 0.37$. $Cr_2Se_3$ bulk samples were reported to be narrow-bandgap semiconductors [1,3, 9-11,13,14,19-22, 57-59]. The large-area continuous nature of the films and the insulating sapphire substrate enabled us to perform temperature-dependent resistivity measurements of the as-grown films using a standard van der Pauw geometry. Figure 5 shows the electrical properties of a 5 nm epitaxial $Cr_2Se_3$ thin film grown on $Al_2O_3$(0001) substrate. Electrical resistivity measured for the temperature range from RT to 77 K, shown in Fig. 5(a), shows semiconducting behavior, *i.e.*, resistance increases with decreasing temperature.

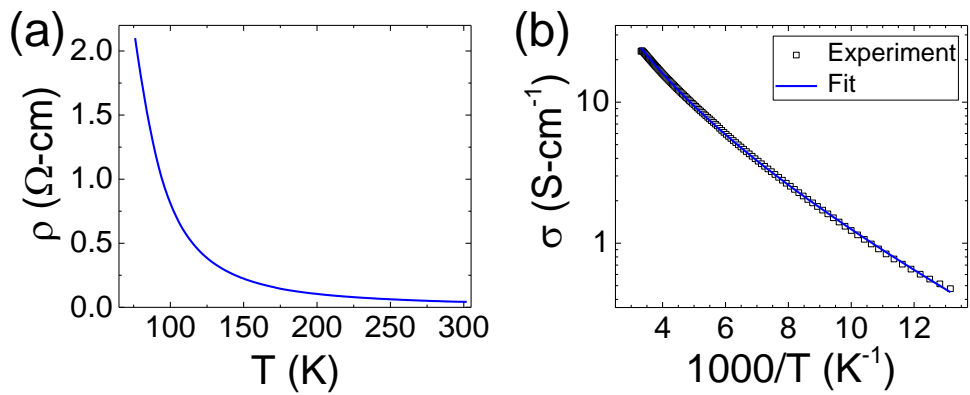

**Figure 5:** Electrical transport properties of the 5 nm epitaxial $Cr_2Se_3$ film grown on $Al_2O_3$(0001) substrate surface. (a) Temperature dependence of electrical resistivity shows an insulating trend. (b) The variation of conductivity is plotted on a semilog scale vs 1/T.

The variation of conductivity (σ) with inverse temperature is shown in Fig. 5(b) on a semilog scale. The low bandgap of Cr$_2$Se$_3$ may indicate an intrinsic like behavior of the material at the measured temperature range [14, 60]. The conductivity of an intrinsic semiconducting material is given by $\sigma = e\, n_i(\mu_e + \mu_h)$, where, $\mu_e$, and $\mu_h$ are carrier mobilities for electrons and holes, respectively, $e$ is the electron charge, and $n_i$ is the intrinsic carrier concentration with $n_i \approx n_0\, T^{\frac{3}{2}} e^{-\frac{E_g}{2k_BT}}$. The factor $T^{\frac{3}{2}}$ is due to the variation of effective density of states with temperature and $e^{-\frac{E_g}{2k_BT}}$ is the Boltzmann weight. For an estimation of the bandgap, the conductivity is fitted with [60]

$$\log \sigma = C + \frac{3}{2} \log T - \frac{E_g}{2k_BT} \qquad (1)$$

where, $E_g$ is the bandgap, $k_B$ is the Boltzmann constant and $C$ is a constant. In Eqn. (1), the carrier mobilities are assumed to be constant with temperature. The value of $E_g$ estimated from the fit is about 0.034 eV. [Assuming temperature dependence of the net mobility $(\mu_e + \mu_h) \propto T^p$ and fitting the data with $\log \sigma = C + \left(\frac{3}{2} + p\right) \log T - \frac{E_g}{2k_BT}$ in a 3-parameter fit instead of the 2-parameter fit of Eqn. (1), we obtain $p \approx -0.12$ and $E_g \approx 0.037$ eV (see Sec. S2 in the Supplemental Material [25]).] In the literature, it was shown that Si and Ge behave intrinsic-like above a temperature $T_m$, where $k_BT_m$ are about 0.04 $E_g$ and 0.08 $E_g$, respectively [60]. The estimated low bandgap of Cr$_2$Se$_3$ is about $5k_BT$ at 77 K (*i.e.*, $k_BT$ is about 0.2 $E_g$) indicating that for the entire temperature range (from RT to 77 K) the material may behave intrinsically, and provides a self-consistent check that Eqn. (1) is valid to describe the temperature dependent resistivity.

### C. Magnetic properties

Previous studies of Cr$_2$Se$_3$ bulk crystals report an antiferromagnetic nature with the Néel temperature ($T_N$) around 42 - 45 K [6-9]. It is to be noted that molecular oxygen trapped in the measurement chamber also undergoes an antiferromagnetic transition at about 43 K and can show similar signature in the magnetic measurement at the same temperature range as in Cr$_2$Se$_3$ [61]. To avoid any possible errors due to trapped oxygen, we have adopted a different approach to investigate the magnetic properties of epitaxial Cr$_2$Se$_3$ thin films. An 8 nm thin layer of ferromagnetic Fe is deposited on top of epitaxial Cr$_2$Se$_3$ layer and capped with a 10 nm Ta layer. A schematic of this exchange-biased structure is shown in Fig. S3(a) in Ref. [25]. MR of the stack is studied to examine the exchange bias phenomenon of this antiferromagnetic-ferromagnetic (AFM-FM) system. Exchange bias effect in the magnetic hysteresis loop is very well known for a FM film coupled with an AFM film where an exchange coupling between the interface spins in AFM and FM gives rise to a shift of the hysteresis loop along the applied magnetic field axis and the magnitude of the shift is defined as the exchange anisotropy field ($H_{EB}$). This exchange anisotropy is observed when the AFM/FM system is cooled in the presence of a static magnetic field (also known as the cooling field, $H_{cool}$) to a temperature below $T_N$. [28,62]. The cooling field aligns all ferromagnetic domains along its direction. The exchange bias phenomenon in AFM/FM system has been extensively studied for its applications in magnetic read heads, magnetic random access memories, high density magnetic recording, *etc*. [63].

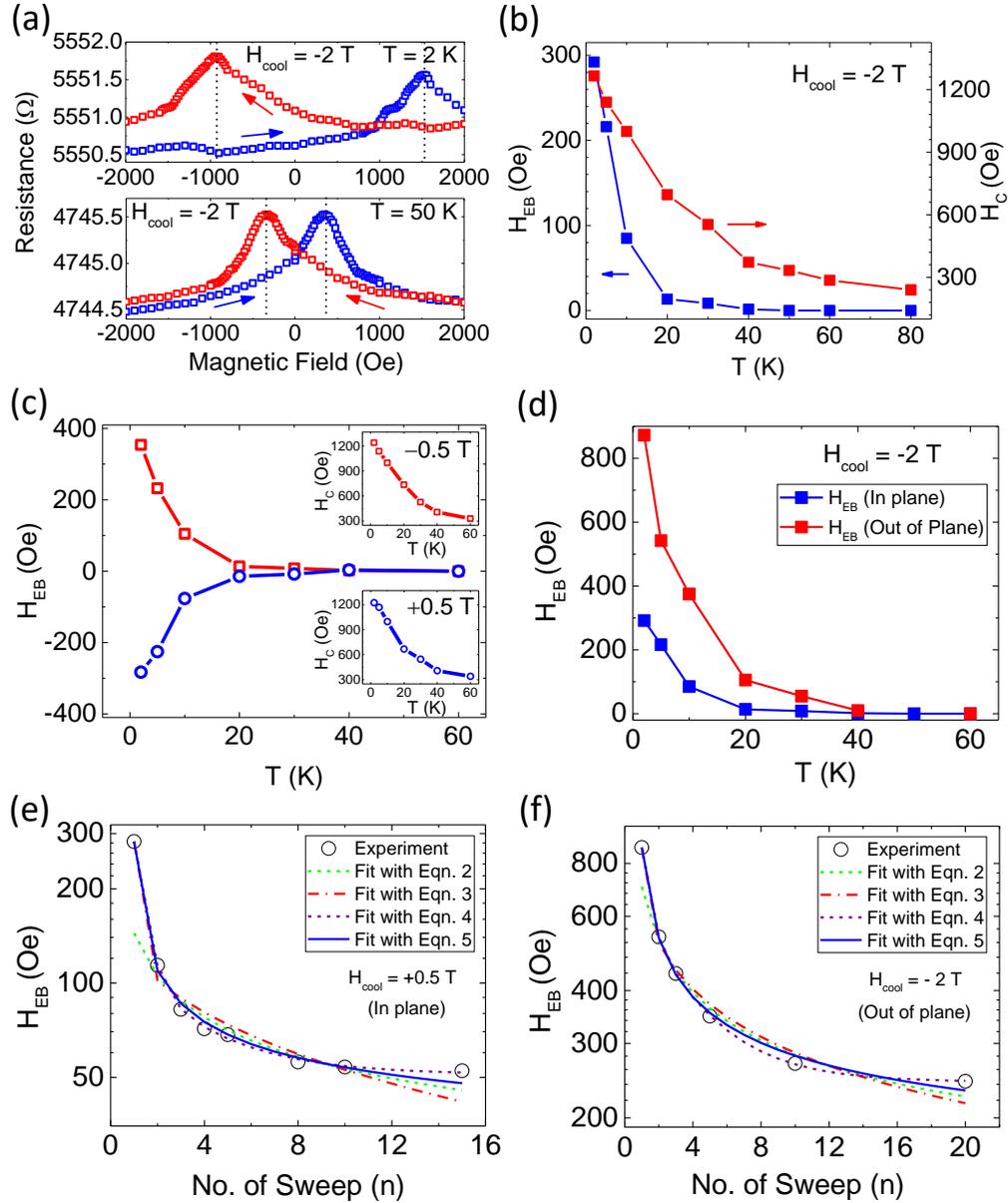

**Figure 6:** Magnetoresistance measurements results of a 25 nm $Cr_2Se_3$ film grown on Si(111)-7×7 surfaces. (a) In-plane magnetic field dependence of the longitudinal magnetoresistance of Ta/Fe/$Cr_2Se_3$/Si(111) at 2 K and 50 K. Vertical dashed lines represent the characteristic values $H_{C1}$ (decreasing field) and $H_{C2}$ (increasing field), for cooling field $H_{cool} = -2$ T. (b) Temperature dependence of in-plane exchange bias field ($H_{EB}$) and coercive field ($H_C$). (c) Temperature dependence of $H_{EB}$ for different cooling fields. Corresponding coercive fields are also shown (inset). (d) Temperature dependence of out-of-plane $H_{EB}$. The in-plane $H_{EB}$ is also plotted for comparison. The cooling field is –2 T for both the measurements. (e) Variation of $H_{EB}$ with number of field cycles (*n*) after in-plane field cooling demonstrating the training effect for the applied field along the surface. The open circles and the lines represent the experimental data points and fit to the experimental data using different models, respectively. (f) Variation of the training for the out-of-plane $H_{EB}$ with number of field cycles (*n*). Experimental data (open circles) are fitted with different models (lines).

MR measurements were performed by the standard Van der Pauw method in a PPMS system capable of applying magnetic fields up to 9 T and the temperatures down to 2 K. The deposited Fe layer of 8 nm thickness is expected to have in-plane magnetization. As shown in Fig. 6(a), with a magnetic field (as well as the cooling field) applied along the sample surface, the MR measurements show typical anisotropic MR (AMR) during field-sweep with two peaks ($H_{C1}$ and $H_{C2}$) corresponding to the coercive field of the Fe layer [64]. Exchange anisotropy field $H_{EB}$ then can be obtained from the two coercive fields as $H_{EB} = (H_{C1} + H_{C2})/2$. To capture the temperature dependence of the $H_{EB}$ with in-plane cooling field, we have repeated MR measurements at different temperatures, each time cooling the sample from RT in presence of the same magnetic field of –2 T along the in-plane direction. Fig. 6(a) shows the variation of MR with respect to magnetic field measured at 2 K and 50 K. The MR hysteresis remains symmetric with respect to $B = 0$ at any temperatures from 300 K down to 50 K, as shown in Fig. 6(a) for 50 K. Asymmetric MR hysteresis appears below 40 K, due to the exchange bias phenomenon and the relative positions of the peaks are displaced against $B = 0$ into the negative magnetic field direction, as shown in Fig. 6(a) at 2 K. MR hysteresis measured on a controlled sample without the $Cr_2Se_3$ layer [Fig. S3(b) in Ref. [25]] remained symmetric for all temperatures (as shown in Fig. S4 in Ref. [25]), confirming that the shift observed in the $Cr_2Se_3$ layer coupled with Fe film is due to an exchange bias from the $Cr_2Se_3$ layer. The obtained exchange bias fields ($H_{EB}$) at different temperatures, for a cooling field, $H_{cool} = -2$ T, are plotted (blue curve) in Fig. 6(b). The curve shows a detectable non-zero $H_{EB}$ at and below 40 K. The temperature above which the exchange bias vanishes is the blocking temperature ($T_B$). It has been reported that thin AFM films with smaller grain sizes often show a lower blocking temperature than the bulk sample Néel temperatures ($T_B < T_N$) [28,65,66]. An equal blocking and Néel temperatures ($T_B \approx T_N$) are observed when the thickness of the AFM layer is increased [28, 66-69]. In our case, the exchange bias effect yields the blocking temperature ($T_B \sim 40$ K) which is quite close to the reported $T_N$ values of bulk $Cr_2Se_3$ samples ($T_N \sim$ 42-45 K) [6-9]. As the temperature is lowered below 40 K, the exchange coupling increases the shift in the peaks and hence, increasing $H_{EB}$ [blue curve in Fig. 6(b)]. With lower temperature, the net magnetization of the AFM layer along the interface induced by the exchange interaction during field cooling also increases [70-74]. Corresponding coercive field defined as $[H_C = |H_{C1} - H_{C2}|/2]$ also increases with decreasing temperature, as shown in Fig. 6(b) (red curve). An enhancement in coercivity below $T_N$ is due to the formation of AFM order in the sample. Variation of the two characteristic peaks ($H_{C1}$ and $H_{C2}$) in AMR measurements with temperature are shown in Fig. S5 in Ref. [25]. Fig. 6(c) shows the variation of $H_{EB}$ with temperature for different $H_{cool}$ values. Reversing the direction of $H_{cool}$ also reverses the sign of $H_{EB}$ but the magnitudes remain relatively unchanged. In general, the shift due to the exchange bias is in the direction opposite to the applied cooling field. Hence, with the change in polarity of $H_{cool}$, corresponding $H_{EB}$ also changes its sign. Corresponding coercivity $H_C$, shown in the inset, also shows the same trend for different cooling fields (also shown in Fig. S6 in Ref. [25]).

Fig. 6(d) shows the variation of $H_{EB}$ with temperature for a cooling field, $H_{cool} = -2$ T, applied along the perpendicular direction to the surface. For comparison, variation of $H_{EB}$ with temperature obtained from in-plane applied fields is also shown in the same figure. It is clearly evident that the magnitude of exchange bias field is higher along the perpendicular to the surface. Corresponding variation of $H_C$, and the characteristic peaks ($H_{C1}$ and $H_{C2}$) are shown in Fig. S7 in Ref. [25]. Previously, both in-plane and the out-of-plane exchange bias have been observed in Co/CoO bilayer film, as well as in different AFM-FM multilayer systems [27,29, 75-77]. For the field cooled samples with the measurement field applied along the easy axis of the FM layer (in-plane), it is expected to have a higher exchange bias. Although the Fe layer is expected to have the easy axis along the surface plane, at all temperatures below 40 K, we observe a higher magnitude of exchange bias for the out-of-plane field orientation. At 2 K, perpendicular $H_{EB}$ is about 3 times higher than that along the surface. This indicates that the exchange bias in our case is highly dependent on the crystalline orientation of the AFM layer that, in turn, may influence the coupling between the AFM and FM layer.

Considering an ideal interface between the AFM and FM layers, an interfacial exchange energy density $\Delta E_{interface}$ of the AFM-FM interface can be estimated as [68]

$$\Delta E_{interface} = |H_{EB}| M_{S(FM)} t_{FM} \qquad (2)$$

where, $|H_{EB}|$ is the magnitude of exchange bias, $M_{S(FM)}$ and $t_{FM}$ are the saturation magnetization and the thickness of the FM layer, respectively. Fig. S8 shows the variation of $\Delta E_{interface}$ at different temperatures for the magnetic field applied along the surface [25]. Considering the FM layer thickness ($t_{FM}$) of 8 nm and $M_{S(FM)} \sim 1420$ emu/cm$^3$, and $|H_{EB}| \sim 292$ Oe at 2 K obtained with in-plane magnetic fields, we estimate $\Delta E_{interface}$ for Cr$_2$Se$_3$/Fe interface is $\sim 0.3$ erg/cm$^2$. For the magnetic field applied perpendicular to the surface, $\Delta E_{interface}$ at 2 K is $\sim 1$ erg/cm$^2$ (as shown in Fig. S9 in Ref. [25]), which is about 3 times higher than that along the plane. The difference in the values of interface energy densities for the two directions perpendicular to each could arise from the preferential orientation of the spins due to crystallinity of the layers [78].

### D. Exchange Bias Training Effect

Both $H_{EB}$ and $H_C$ tend to decrease monotonically when the AMR measurement is repeated multiple times at the same temperature after the initial field cooling, which is known as the training effect [31]. Absence of a net magnetic moment in an AFM produces no net Zeeman energy in an external magnetic field, which results in randomly oriented domains during the first field cooling from above the Neel temperature. Multiple cycling of the hysteresis loop gradually rearranges the spin structure of the AFM layer relaxing it towards its ground state giving the observed training effect [35,79]. The strength of the training effect depends mostly on the exchange interaction at the interface, change of non-equilibrium spin moment of the AFM domains and on the thermal energy. Fig. 6(e) shows the dependence of in-plane $H_{EB}$ on the loop number ($n$) measured for 15 consecutive cycles at 2 K after initial field cooling with $H_{cool} = 0.5$ T. Corresponding in-plane $H_C$, in Fig. S10, also shows a decreasing trend with the number of sweeps [25]. The drop in $H_{EB}$ is maximum only after the first cycle (down to $\sim 40\%$), as shown in the inset in Fig. S10 in Ref. [25]. The variation is less for the subsequent cycles ($\sim 20\%$, only) and stabilizes towards a constant level, all of which point towards some underlying relaxation dynamics at the AFM-FM interface. Although the microscopic origin is still debatable, different theoretical models were proposed to explain the training effect of $H_{EB}$ based on the time dependence of the interface spin moment of the AFM layer. We attempt to analyze the observed training effect by fitting our experimental data with different models.

We first follow the thermal relaxation model, where the dependence of $H_{EB}$ and the number of cycles ($n$) follows a simple empirical power-law [31]:

$$H_{EB}(n) = H_{EB}(\infty) + \frac{k_H}{\sqrt{n}} \qquad (3)$$

where $H_{EB}(n)$ and $H_{EB}(\infty)$ are the exchange bias fields at the $n^{th}$ cycle, and in the limit of infinite cycles ($n \to \infty$), respectively, and $k_H$ is a system-dependent constant. The fit result with Eqn. (3), as shown in Fig. 6(e) (green dashed line), is in good agreement with experimental data for $n > 1$, which is consistent with previous studies in the literature [35,68]. The fit breaks down and results in a negative $H_{EB}(\infty)$ if the data point at $n = 1$ is included. Excluding the data at $n = 1$, the values of $H_{EB}(\infty)$ and $k_H$ obtained are 11.2 Oe and 133.2 Oe, respectively. To explain the variation of $H_{EB}$ including $n = 1$, Binek [35] derived the following relation for the exchange bias training effect:

$$H_{EB}(n+1) = H_{EB}(n) + \gamma_H [H_{EB}(n) - H_{EB}(\infty)]^3 \qquad (4)$$

where $\gamma_H$ is a system-dependent constant. Eqn. (3) recovers the empirical power-law Eqn. (2) in the limit of $n \gg 1$ with $2\gamma_H = 1/k_H^2$ [35]. However, the fit using Eqn. (3) including $n = 1$ data is not satisfactory in our case either, unless a negative $H_{EB}(\infty)$ is allowed [red dashed line in Fig. 6(e)]. Excluding $n = 1$ data point allows a better fit to the data (shown in Fig. S11 in Ref. [25]) giving $H_{EB}(\infty) = 23.8$ Oe and $\gamma_H$ about $445.6 \times 10^{-7}$ Oe$^{-2}$ corresponding to $k_H = 105.9$ Oe.

To allow for a positive $H_{EB}(\infty)$ and explain the change of $H_{EB}$ including the $n = 1$ data, we consider an alternative model [46,47] that describes the training effect using exponential time-dependence of a mixed state of frozen and rotatable uncompensated spins at the interface:

$$H_{EB}(n) = H_{EB}(\infty) + A_f \exp\left(-\frac{n}{P_f}\right) + A_i \exp\left(-\frac{n}{P_i}\right) \qquad (5)$$

where, $A_f$ and $P_f$ are parameters related to the changes of the frozen spins and $A_i$ and $P_i$ are parameters for the rotatable spin component at the AFM-FM interface. Here, dimensionless parameters $P_f$ and $P_i$ act as the relaxation time constants for the exponential decay of the spin components towards equilibrium. The above equation fits the experimental data very well [purple dashed line in Fig. 6(e)] indicating a complex spin arrangement at the interface. The parameter values obtained from the fit are, $H_{EB}(\infty) = 51.2$ Oe, $A_f = 71.9$ Oe, $P_f = 3.2$, $A_i = 1333.2$, and $P_i = 0.5$. The associated larger pre-factor $A_i \gg A_f$ and the relaxation time ratio $P_f/P_i \sim 6.48$ indicates that the exchange bias is governed mainly by the interface spins. However, this model predicts a much higher $H_{EB}(\infty) = 51.2$ Oe, compared to Eqn. (3) and Eqn. (4) (excluding $n = 1$ data).

To explain our data with the same physical model of Binek [35] and to allow for a positive $H_{EB}(\infty)$ including the $n = 1$ data, we next consider a modified power law model [38,39]:

$$H_{EB}(n) = H_{EB}(\infty) + k_H[n + n_0]^{-0.5} \qquad (6)$$

where, $k_H$ is the same constant as in Eqn. (3), and $n_0$ is a dimensionless number. It should be noted that both Eqns. (3) and (4) can be derived from Eqn. (6) under certain approximations (see Sec. S12 in the Supplemental Material [25]). $H_{EB}$ obtained from Eqn. (6) allows a much better fit to our experimental data

[blue solid line in Fig. 6(e)] including the $n = 1$ data point compared to Eqn. (3) or (4), while preserving the power law dependence. The parameter values obtained from the fit are: $H_{EB}(\infty) = 23.7$ Oe, $k_H = 90.9$ Oe.

The training effect along the out-of-plane direction is shown in Fig. 6(f). Again, a decreasing trend of $H_{EB}$ on the number of sweep ($n$) measured for 20 consecutive cycles, at an initial $H_{cool}$ of $-2$ T and a constant temperature of 2 K, is observed. Corresponding $H_C$ also shows a declining trend with the number of sweeps (as shown in Fig. S13 in Ref. [25]). Fitting the out-of-plane exchange bias training effect for $n > 1$ with Eqn. (3) gives $H_{EB}(\infty) = 86.2$ Oe and $k_H = 618.5$ Oe. Fit using Eqn. (4) including $n = 1$, again results in a much lower $H_{EB}(\infty) = 22.8$ Oe, which improves with the exclusion of the data at $n = 1$ ($H_{EB}(\infty) = 99.0$ Oe, as shown in Fig. S14 in Ref. [25]). The exponential dependence of Eqn. (5) explains the data well including $n = 1$ with parameters obtained as, $H_{EB}(\infty) = 243.4$ Oe, $A_f = 457.6$ Oe, $P_f = 3.4$, $A_i = 2168.1$, and $P_i = 0.5$. Similar to the in-plane case, out-of-plane $H_{EB}(\infty)$ predicted from Eqn. (5) is much higher than that obtained from Eqn. (3) or (4) (excluding $n = 1$ data). The relaxation times of frozen and rotatable spins obtained for out-of-plane training effect are quite close to those obtained from the in-plane training effect. Finally, we fit the data with power law of Eqn. (6) [Fig. 6(f) blue solid line]. The fit could explain the data very well including the $n = 1$ data point, with parameters obtained as: $H_{EB}(\infty) = 120.3$, $k_H = 492.3$ Oe. In the Supplemental Material [25] sections S15 and S16, we provide a complete summary of the fitting results of the exchange bias training effect using different models.

Origin of exchange bias is a complex phenomenon that depends on several factors, *e.g.*, interfacial coupling strength, AFM anisotropy, interface domain structure, film thickness, strain, atomic steps, interface roughness [78]. In addition, exchange bias in epitaxial samples is influenced by the field cooling direction and intrinsic magnetocrystalline anisotropy of the AFM layer [80-82]. In our case, AFM layer of $Cr_2Se_3$ is epitaxially grown along the *c*-axis on sapphire substrate, whereas the top Fe film is polycrystalline. Hence, the spins in AFM layer may have possible magnetic easy axis along the out-of-plane direction [49], and could show enhanced exchange anisotropy when field cooled with a magnetic field applied perpendicular to the surface. In addition, depending on the directions, the magnetic field interacts differently with the uncompensated spins (which cannot be ruled out) along with the compensated, and can produce a complex picture in the experiment. A bilayer of $Cr_2Se_3$ layer coupled with another (001)-oriented ferromagnetic thin film with a hexagonal crystal structure and a perpendicular magnetic anisotropy (*e.g.*, $Cr_2Te_3$ thin film, Ref. [49]) would be an interesting out-of-plane exchange bias system to study. However, to obtain a deeper understanding of the exchange bias and the training effect in epitaxial $Cr_2Se_3$ thin films of different thicknesses, further theoretical and experimental studies on its time dependence as well as the spin structure of the AFM-FM system close to the interface should be done in detail.

### IV.    Summary and conclusion

In conclusion, we have carried out the MBE growth of $Cr_2Se_3$ thin films on insulating *c*-$Al_2O_3$(0001) and Si(111)-(7×7) substrates. Structural, electrical and magnetic properties of the films have been characterized by several *in situ* and *ex situ* techniques. Sharp streaks in RHEED patterns imply smooth thin film growth on both the substrates. The film has hexagonal structure and oriented along the (001)- direction (*c*-axis), as confirmed from *in situ* RHEED and STM, and *ex situ* XRD. Chemical composition of the film is investigated through *in situ* XPS measurement. Electrical measurement on the as-grown film shows a narrow bandgap semiconducting behavior. Antiferromagnetic nature of the grown film is confirmed from the magnetotransport measurements of an exchange coupled system of $Cr_2Se_3$(AFM)-Fe(FM). Exchange bias is higher in magnitude along the out-of-plane direction compared to that in the in-plane direction. Exchange bias training effect in both directions seems to be consistent with a modified power-law decay behavior. Our results indicate that epitaxial $Cr_2Se_3$ thin films could offer an interesting material system to study effects of magnetic anisotropy and field cooling direction on the exchange bias properties in fully epitaxial AFM-FM bilayers.

**Conflict of interest:** The authors declare no competing financial interest.

**Acknowledgment:** This work was supported in part by an NSF EFRI grant, NSF NASCENT ERC and NSF NNCI (done at the Texas Nanofabrication Facility at the University of Texas at Austin supported by NSF grant NNCI-1542159). We appreciate technical support from Omicron.

# Supplemental Material

# Structural and Magnetic Properties of Molecular Beam Epitaxy Grown Chromium Selenide Thin Films

Anupam Roy [1]*, Rik Dey [1], Tanmoy Pramanik [1], Amritesh Rai [1], Ryan Schalip [1], Sarmita Majumder [1], Samaresh Guchhait [2], and Sanjay K Banerjee [1]

[1] Microelectronics Research Center, The University of Texas at Austin, Austin, Texas 78758, USA
[2] Department of Physics and Astronomy, Howard University, Washington, DC 20059, USA

*Address correspondence to: anupam@austin.utexas.edu.

**Contents**

**S1:** RHEED from $Cr_2Se_3$ thin films of different thicknesses
**S2:** Estimation of bandgap assuming temperature-dependent mobility
**S3:** Schematic of exchange bias system and controlled sample
**S4:** Exchange bias in a controlled sample
**S5:** Temperature dependence of characteristic peaks ($H_{C1}$ and $H_{C2}$) with in-plane field
**S6:** Temperature dependence of exchange bias and coercive field for varying cooling fields
**S7:** Temperature dependence of characteristic peaks ($H_{C1}$ and $H_{C2}$) with out-of-plane field
**S8:** Temperature variation of in-plane interface energy per unit area
**S9:** Temperature variation of out-of-plane interface energy per unit area
**S10:** In-plane training effect
**S11:** Fit to the in-plane training effect using Eqn. (4) excluding $n = 1$
**S12:** Derivation of the modified power law equation (Eqn. 6)
**S13:** Out-of-plane training effect
**S14:** Fit to the out-of-plane training effect using Eqn. (4) excluding $n = 1$
**S15:** Parameters obtained from the fit to the exchange bias training effect
**S16:** Validity of training effect fit with Eqn. (4)

## S1: RHEED from Cr$_2$Se$_3$ thin films of different thicknesses

Figure S1 shows the RHEED patterns before and after Cr$_2$Se$_3$ growth on Al$_2$O$_3$(0001) and Si(111)-(7×7) substrates of different thicknesses.

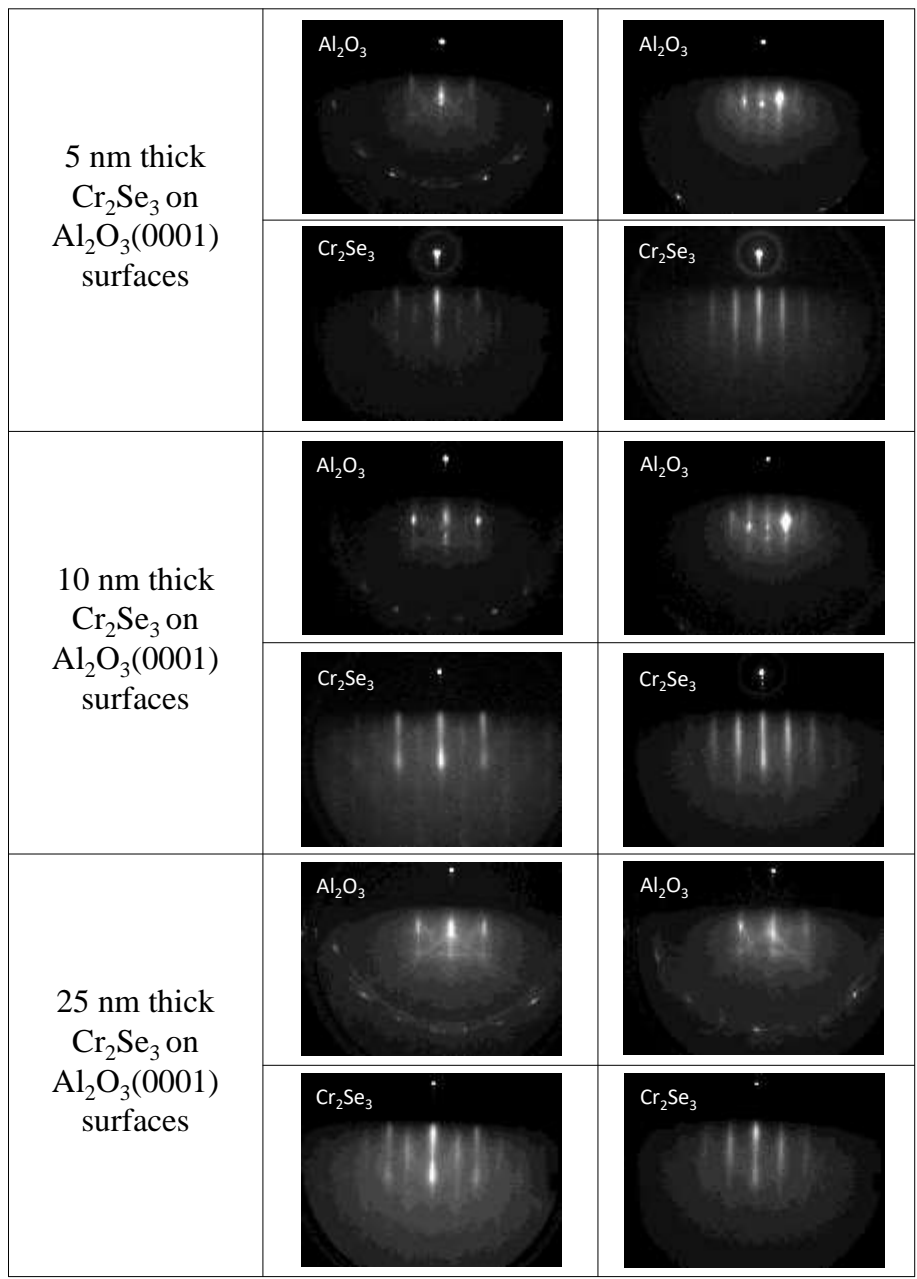

**Figure S1 (a):** RHEED patterns following Cr$_2$Se$_3$ growth on Al$_2$O$_3$(0001) surfaces. Left and right panels correspond to the patterns along [1 0 -1 0] and [1 1 -2 0] orientations of Al$_2$O$_3$, respectively.

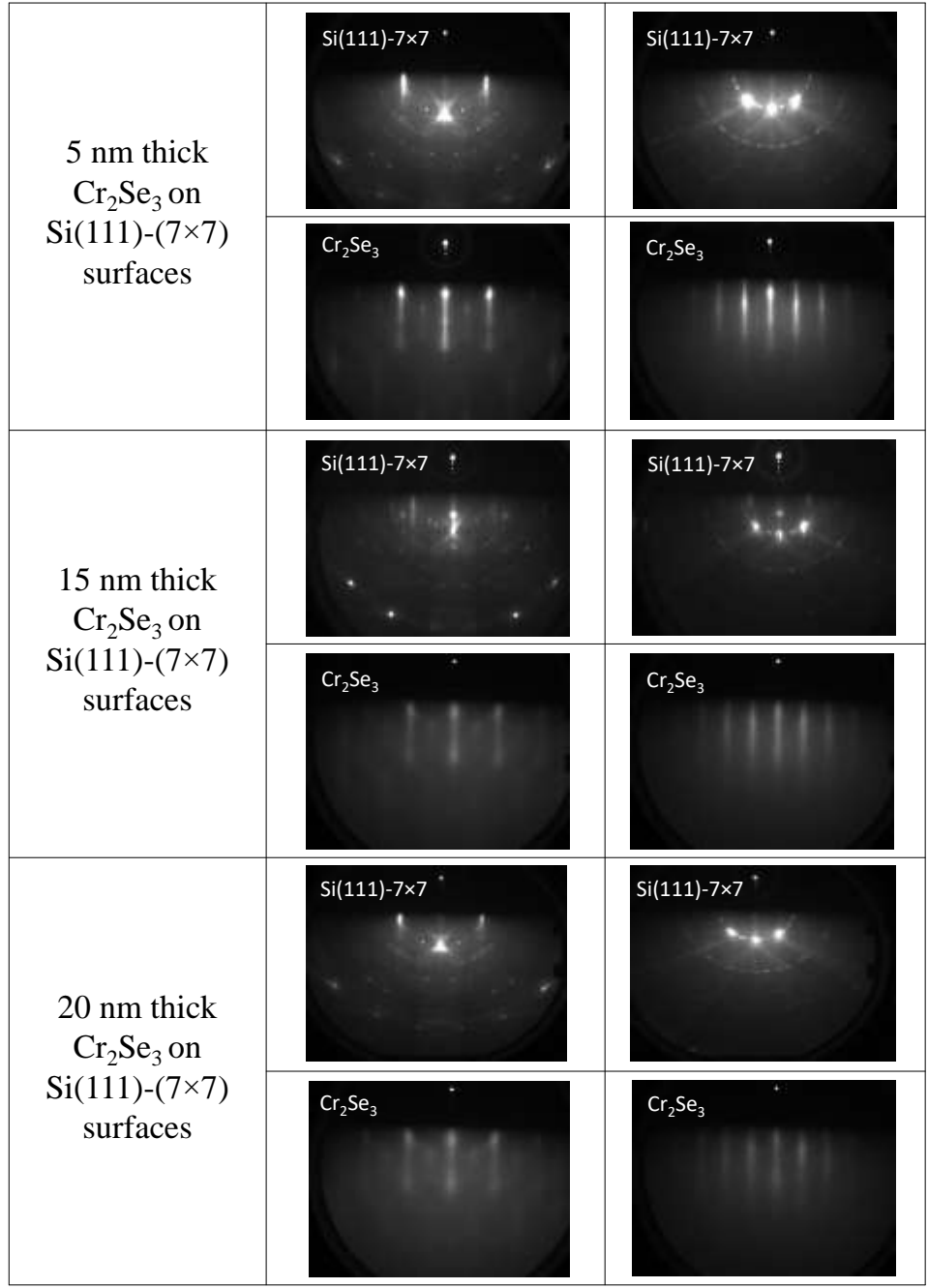

**Figure S1 (b):** RHEED patterns following $Cr_2Se_3$ growth on Si(111)-(7×7) surfaces. Left and right panels correspond to the patterns along [1 1 -2] and [1 -1 0] orientations of Si, respectively.

## S2: Estimation of bandgap assuming temperature-dependent mobility

Considering the mobility ($\mu$) depending on the temperature ($T$), we assume the net mobility ($\mu_e + \mu_h$) $\propto T^p$ and fit the data with

$$\log \sigma = C + \left(\frac{3}{2} + p\right) \log T - \frac{E_g}{2k_B T} \qquad (S1)$$

This is a 3-parameter fit instead of the 2-parameter fit of Eqn. (1) in the main text. From the fit we obtain $p \approx -0.12$ and $E_g \approx 0.037$ eV.

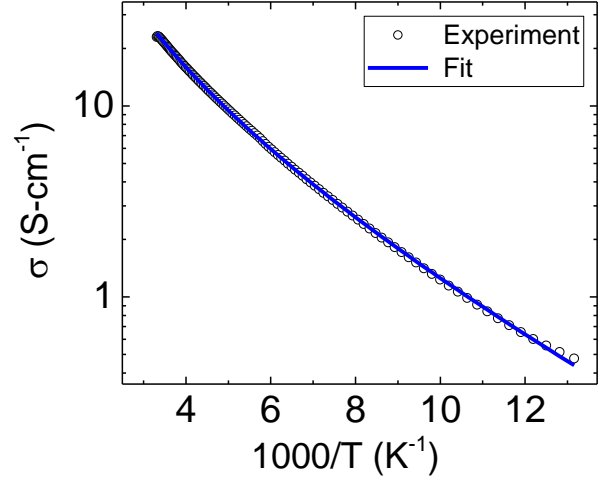

**Figure S2:** The variation of conductivity on a semilog scale vs 1/T.

### S3: Schematic of exchange bias system and controlled system

Figure S3 shows the schematic of the exchange bias system. An antiferromagnetic $Cr_2Se_3$ layer (25 nm) and a ferromagnetic Fe layer (8 nm) are coupled and capped with 10 nm Ta layer [Fig. S3 (a)]. A similar schematic of the controlled sample without the antiferromagnetic $Cr_2Se_3$ layer is shown in Fig. S3 (b). The thicknesses of Fe and Ta layers in the controlled sample are same as in the exchange-biased sample in Fig. S3 (a).

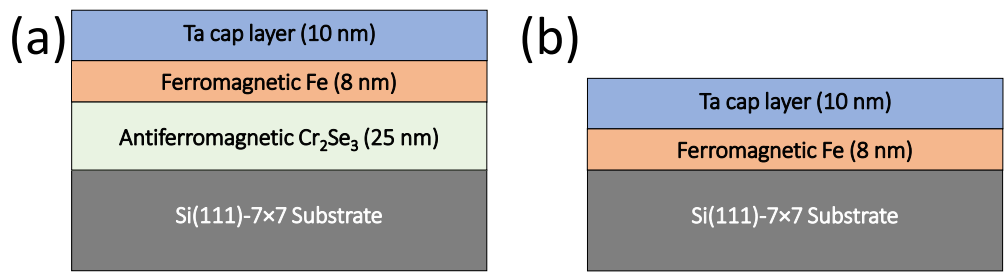

**Figure S3:** Schematic of (a) the AFM/FM exchange bias system and (b) a controlled sample without the antiferromagnet $Cr_2Se_3$ layer.

### S4: Exchange bias in a controlled sample

Figure S4 shows magnetoresistance (MR) measurement performed at 5 K on a controlled Ta/Fe/Si(111) sample with in-plane magnetic field. Peaks in MR (vertical dotted lines) are symmetric with respect to $B = 0$ confirming no exchange bias in this material system. This confirms that the shift observed in MR in Fig. 6(a) are due to an exchange bias in AFM/FM system.

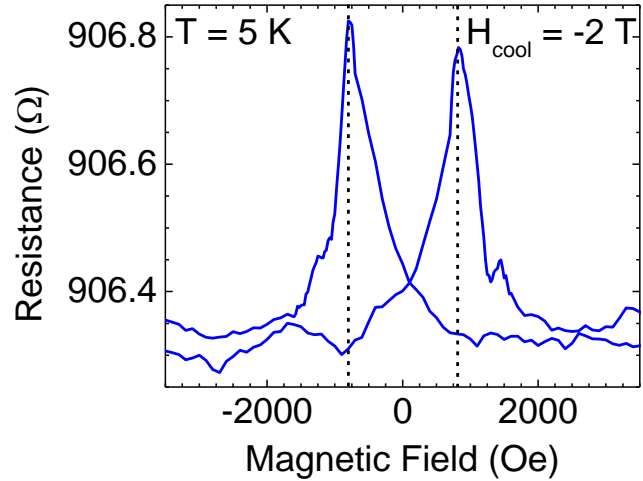

**Figure S4:** In-plane magnetic field dependence of the longitudinal magnetoresistance of Ta/Fe/Si(111) controlled sample at 5 K shows no exchange bias. The sample is cooled down from RT to 5 K with the in-plane cooling field ($H_{cool}$) of –2 T.

### S5: Temperature dependence of characteristic peaks ($H_{C1}$ and $H_{C2}$) with in-plane field

Variations of two characteristic peaks ($H_{C1}$ and $H_{C2}$) in AMR measurements with temperature are shown in Fig. S5. Magnetic fields applied during the measurement, as well as during the field cooling are along the surface. Corresponding variations of exchange biases and the coercive fields with temperature are also shown. For each measurement at different temperatures, the sample was cooled from RT down to the measurement temperature in presence of a magnetic field of –2 T. Both $H_{C1}$ and $H_{C2}$ are seen to be decreasing with increasing temperature.

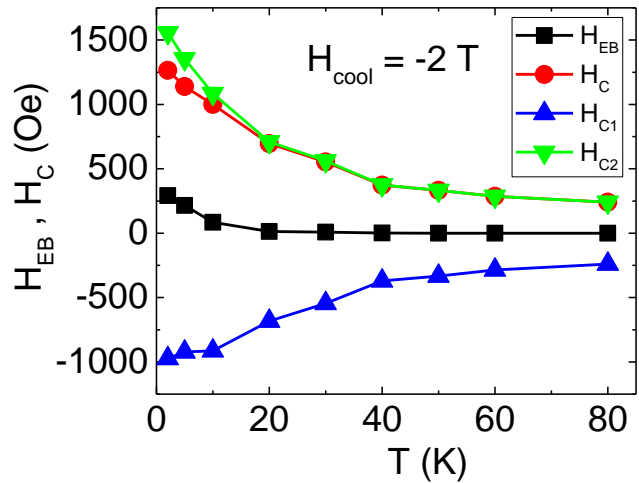

**Figure S5:** Temperature dependence of in-plane exchange bias field ($H_{EB}$), coercive field ($H_C$), and the characteristic fields $H_{C1}$ and $H_{C2}$. The cooling field is –2 T.

## S6: Temperature dependence of exchange bias and coercive field for varying cooling fields

In-plane exchange bias field ($H_{EB}$) and coercive field ($H_C$) variations with temperature for different cooling fields ($H_{cool}$) are shown in Fig. S6. In all cases, $H_C$ magnitude increases with decreasing temperature, a typical characteristic of a ferromagnetic film [26]. The nature of $H_C$ vs. $T$ and the magnitude of $H_C$ are almost insensitive to the magnitudes of cooling fields.

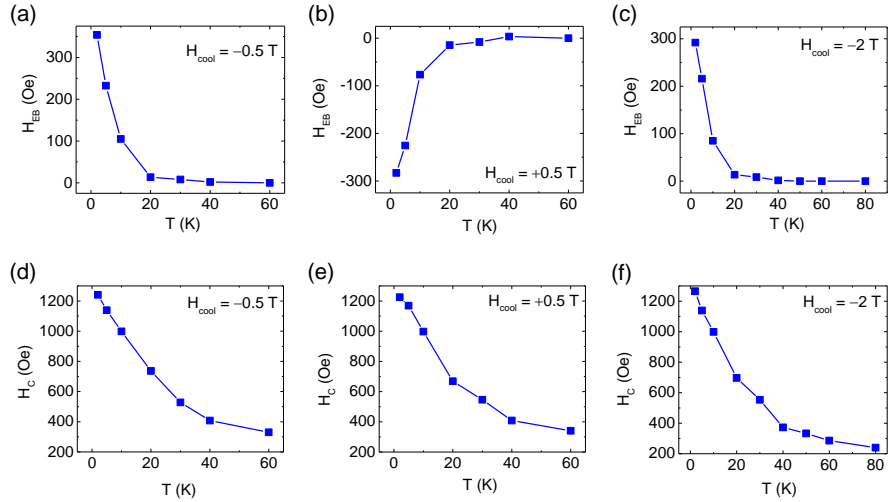

**Figure S6:** Temperature dependence of the in-plane exchange bias field ($H_{EB}$) and coercive field ($H_C$) for different cooling fields.

## S7: Temperature dependence of characteristic peaks ($H_{C1}$ and $H_{C2}$) with out-of-plane field

Similar to the in-plane variations (Fig. S5), temperature dependence of the characteristic peaks ($H_{C1}$ and $H_{C2}$) with in perpendicular magnetic field measurements are shown in Fig. S7. Corresponding variations of exchange biases and the coercive fields with temperature are also shown. The nature matches exactly with the in-plane measurement in Fig. S5 – magnitudes of both $H_{C1}$ and $H_{C2}$ are decreasing with increasing temperature.

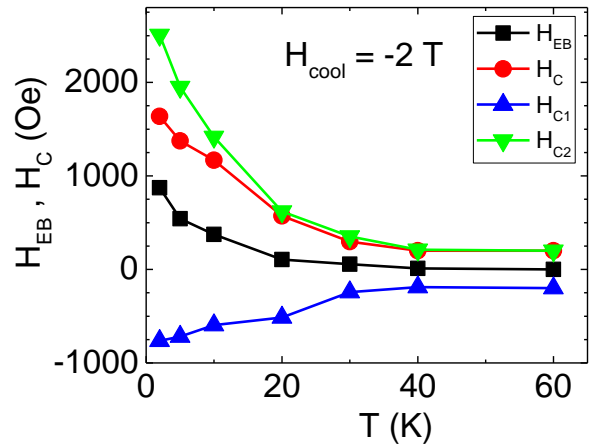

**Figure S7:** Temperature dependence of out-of-plane exchange bias field ($H_{EB}$), coercive field ($H_C$), and the characteristic fields $H_{C1}$ and $H_{C2}$. The cooling field is –2 T.

## S8: Temperature variation of in-plane interface energy per unit area

Figure S8 shows the variation of interface energy per unit area with temperature for the magnetic field applied along the surface. The variation of interface energy per unit area, as estimated from Eqn. (2), $\Delta E_{interface} = |H_{EB}| M_{S(FM)} t_{FM}]$, shows a decreasing trend with increasing temperature for different cooling fields.

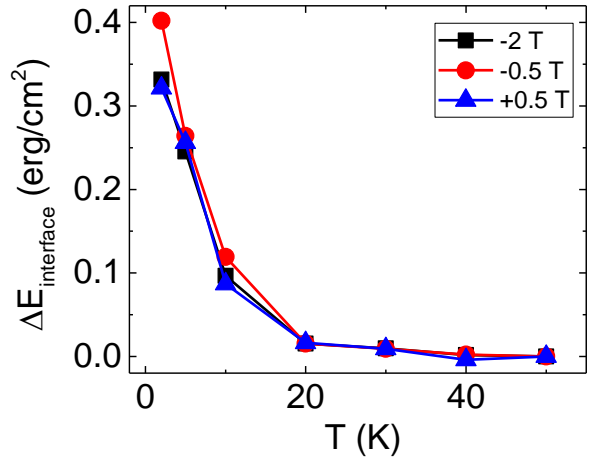

**Figure S8:** Variation of the interface energy per unit area with temperature for the in-plane magnetic field measurement.

## S9: Temperature variation of out-of-plane interface energy per unit area

Figure S9 shows the variation of interface energy per unit area with temperature for the magnetic field applied normal to the surface. Using Eqn. (2), the estimated value of $\Delta E_{interface}$ at 2 K, for the magnetic field (–2 T) applied perpendicular to the surface, is about 1 erg/cm$^2$. The variation of the same for the magnetic field applied along the surface is also plotted for comparison. Along the surface, the estimated value of $\Delta E_{interface}$ at 2 K is about 0.3 erg/cm$^2$. Similar observation has been made for different AFM-FM core-shell and multilayer structures [27-29].

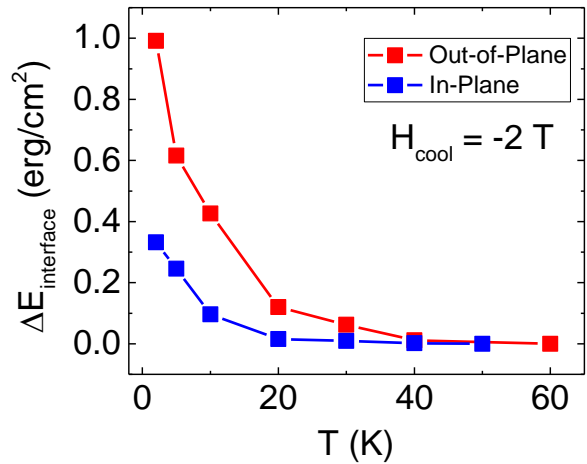

**Figure S9:** Variation of the interface energy per unit area with temperature for the out-of-plane and in-plane magnetic field measurement.

## S10: In-plane training effect

Figure S10 shows the in-plane training of the exchange bias field at 2 K, with the cooling field applied along the surface. Because of the training effect, a monotonic decrease of the exchange bias field when cycling the magnetic field through consecutive sweeps is seen in both cases. Corresponding coercive field also decreases gradually with increasing $n$.

Inset of Fig. S10 shows the percentage change of training as calculated from the equation below [30]:

$$TE_n(\%) = \left[1 - \frac{H_{EB}^1 - H_{EB}^n}{H_{EB}^1}\right] \times 100 \ (\%) \tag{S2}$$

As observed, the in-plane drop in $H_{EB}$ is about 40 % after the first cycle, and the drop rate becomes more subtle for the following cycles.

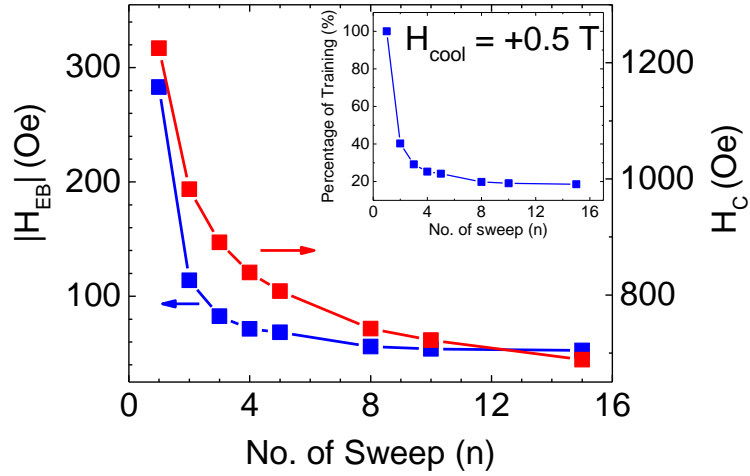

**Figure S10:** Variation of $H_{EB}$ and $H_C$ with number of field cycles ($n$) after field cooling demonstrating the training effect for the applied field along the surface. Inset shows the percentage change in the in-plane training with $n$.

## S11: Fit to the in-plane training effect using Eqn. (4) excluding $n = 1$

Fit with Eqn. (3) [31] gives a negative value of $H_{EB}$ ($\infty$) indicating a positive loop shift, which is unexpected from the experimental observations up to $n = 15$. Different training behavior from $n = 1$ to $n = 2$ has been previously explained as a combination of thermal and athermal effects [32,33]. The initial larger change in the coercive field also has been attributed to an irreversible change in the AFM spin moments triggered by the first magnetization reversal of the exchange coupled FM followed by the field cooling [34]. Eqn. (4) [35] is based on Landau-Khalatnikov equation [36,37] for the relaxation of the interface AFM spin and can explain the variation of $H_{EB}$ including $n = 1$ in different magnetic systems. However, for our experimental data a negative $H_{EB}$ ($\infty$) is observed. Similar observations of negative $H_{EB}$ ($\infty$) were also reported in [33]. These values are quite close to the value obtained from Eqn. (3). A better fit to our experimental data, using Eqn. (4), is allowed only with exclusion of $n = 1$ data point. The obtained value of

$H_{EB}(\infty)$ from the fit is about 23.8 Oe and $\gamma_H$ about $445.6 \times 10^{-7}$ Oe$^{-2}$ corresponding to $k_H = 105.9$ Oe. The fitting parameters are given in Table 1 in supplementary information S15, and that matches very well with that using Eqn. (6).

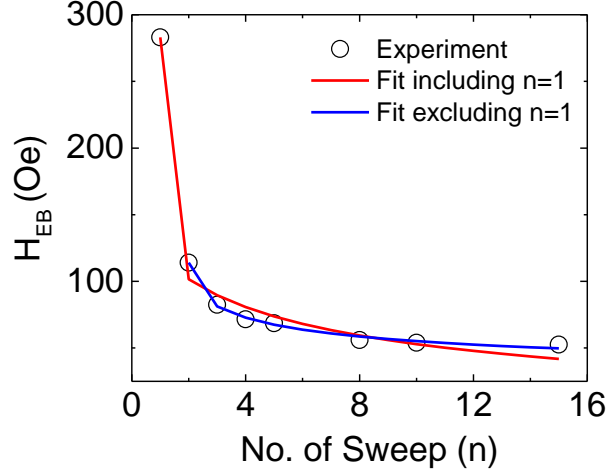

**Figure S11:** Fit (solid line) to the in-plane training effect experimental data (solid square) using Eqn. 4 excluding the data point at $n = 1$. The cooling field is +0.5 T.

**S12: Derivation of the modified power law equation (Eqn. 6)**

To explain our data with the same physical model of Binek [35] as well as to allow for a positive $H_{EB}(\infty)$ and inclusion of the $n = 1$ data, we next consider a modified power law model. This model is based on the relaxation of the interface spin according to Landau-Khalatnikov equation [36,37] and follows the approach of Rui *et al*. [38], which does not invoke the discretization approximation of the spin relaxation with time of Binek [35]. Based on the relaxation of the interface AFM spin the Landau-Khalatnikov Eqn. reads [35-37]:

$$\xi \frac{dS(t)}{dt} = -\frac{\partial \Delta F}{\partial S}, \text{ where } \Delta F = \frac{b}{4}(\delta S(t))^4, \text{ and } \delta S(t) = S(t) - S_e.$$

So, from the above equations we have, $\xi \frac{dS(t)}{dt} = -b\,(\delta S(t))^3$. (S3)

Binek's model [35] is based on the assumption that $S(t)$ does not change between two hysteresis loops, and only in the $n^{th}$ loop $S(t)$ changes from $S_n$ to $S_{n+1}$. So, $\delta S(t)$ changes from $\delta S_n$ to $\delta S_{n+1}$, where $\delta S_n = S_n - S_e$ (where $S_e$ denotes the equilibrium AFM interface magnetization). Also, it is assumed [35] that after large number of cycles, $S(t)$ saturates to the equilibrium value, *i.e.*, $S_\infty = S_e$.

Integrating Eqn. (S3) (exactly) during the $n^{th}$ loop gives

$-\int_{\delta S_n}^{\delta S_{n+1}} \frac{d(\delta S(t))}{(\delta S(t))^3} = \frac{b}{\xi} \int_{t_n}^{t_{n+1}} dt$

$\Rightarrow \frac{1}{(\delta S_{n+1})^2} - \frac{1}{(\delta S_n)^2} = \frac{2b}{\xi}(t_{n+1} - t_n) = \frac{2b\tau}{\xi}$ (S4)

where $\tau = (t_{n+1} - t_n)$ is the measurement time for the $n^{th}$ loop. From Eqn. (S4) we get

$$\frac{1}{(S_{n+1}-S_\infty)^2} - \frac{1}{(S_n-S_\infty)^2} = \frac{2b\tau}{\xi} \tag{S5}$$

Assuming $\tau$ to be the same for each loop, we have

$$\frac{1}{(S_2 - S_\infty)^2} - \frac{1}{(S_1 - S_\infty)^2} = \frac{2b\tau}{\xi}$$

$$\frac{1}{(S_3 - S_\infty)^2} - \frac{1}{(S_2 - S_\infty)^2} = \frac{2b\tau}{\xi}$$

$$\ldots$$

$$\frac{1}{(S_{n+1} - S_\infty)^2} - \frac{1}{(S_n - S_\infty)^2} = \frac{2b\tau}{\xi}$$

Adding the above equations gives

$$\frac{1}{(S_{n+1}-S_\infty)^2} = \frac{1}{(S_1-S_\infty)^2} + \frac{2b\tau}{\xi} n. \tag{S6}$$

Now, using $H_{EB}(n) = kS_n$ (with $H_{EB}(n)$ in Oe) in Eqn. (S6) gives

$$\frac{1}{\left(H_{EB}(n+1)-H_{EB}(\infty)\right)^2} = \frac{1}{\left(H_{EB}(n)-H_{EB}(\infty)\right)^2} + 2\gamma_H, \tag{S7}$$

Here $\gamma_H$ is a system-dependent constant defined as where $\gamma_H = \frac{b\tau}{k^2\xi} = \frac{b}{k^2\tilde{\xi}}$, and $\tilde{\xi} = \frac{\xi}{\tau}$, where $b$ is a constant, $\tau$ is the measurement time, $\xi$ is the damping constant (considered as the inverse relaxation time) and $k$ is proportional to the exchange coupling constant between the AFM/FM interface (as defined by Binek [35]).

Eqn. (S7) reads

$$\left(H_{EB}(n+1) - H_{EB}(\infty)\right) = \frac{\left(H_{EB}(n) - H_{EB}(\infty)\right)}{\sqrt{1 + 2\gamma_H\left(H_{EB}(n) - H_{EB}(\infty)\right)^2}}$$

$$\Rightarrow H_{EB}(n+1) = H_{EB}(n) - \left(H_{EB}(n) - H_{EB}(\infty)\right)\left[1 - \frac{1}{\sqrt{1+2\gamma_H\left(H_{EB}(n)-H_{EB}(\infty)\right)^2}}\right]. \tag{S8}$$

In the limit, $2\gamma_H\left(H_{EB}(n) - H_{EB}(\infty)\right)^2 \ll 1$, expanding the square root in Eqn. (S8) gives

$$H_{EB}(n+1) \approx H_{EB}(n) - \left(H_{EB}(n) - H_{EB}(\infty)\right)\left[1 - \left(1 - \frac{1}{2}2\gamma_H\left(H_{EB}(n) - H_{EB}(\infty)\right)^2\right)\right]$$

$$\Rightarrow H_{EB}(n+1) \approx H_{EB}(n) - \gamma_H\left(H_{EB}(n) - H_{EB}(\infty)\right)^3. \tag{S9}$$

Eqn. (S9) is same as in Binek's model (obtained by discretizing the time derivative [35]).

Now, using $H_{EB}(n) = kS_n$ in Eqn. (S5) gives

$$\frac{1}{\left(H_{EB}(n+1) - H_{EB}(\infty)\right)^2} = \frac{1}{\left(H_{EB}(1) - H_{EB}(\infty)\right)^2} + 2\gamma_H n$$

$$\Rightarrow \left(H_{EB}(n+1) - H_{EB}(\infty)\right) = \frac{\left(H_{EB}(1) - H_{EB}(\infty)\right)}{\sqrt{1 + 2\gamma_H n \left(H_{EB}(1) - H_{EB}(\infty)\right)^2}}$$

$$\Rightarrow H_{EB}(n+1) = H_{EB}(\infty) + \frac{1}{\sqrt{2\gamma_H}} \frac{1}{\sqrt{n + \left\{\sqrt{2\gamma_H}\left(H_{EB}(1) - H_{EB}(\infty)\right)\right\}^{-2}}}$$

$$\Rightarrow H_{EB}(n) = H_{EB}(\infty) + \sqrt{\frac{k^2 \xi}{2b\tau}} \frac{1}{\sqrt{n+n_0}}. \qquad (S10)$$

where $n_0 = \left\{\sqrt{2\gamma_H}\left(H_{EB}(1) - H_{EB}(\infty)\right)\right\}^{-2} - 1$.

Eqn. (S10) is of the same form as derived by Rui *et al* [38].

Also, for large *n*, Eqn. (S10) gives the following Paccard's formula [31],

$$H_{EB}(n) = H_{EB}(\infty) + \frac{k_H}{\sqrt{n}}, \qquad (S11)$$

where $k_H = \frac{1}{\sqrt{2\gamma_H}} = \sqrt{\frac{k^2 \xi}{2b\tau}} = k\sqrt{\frac{\tilde{\xi}}{2b}}$, as derived by Binek [35] from Eqn. (S9).

Equation (6) also was derived by Su and Hu [39] based on spin dynamics simulation employing Landau-Lifshitz-Gilbert equation and assuming a power-law energy dissipation of the AFM layer as the AFM spins move towards equilibrium in an attempt to describe the power law behavior of Paccard *et al.* [31]. This modified power-law model has also been used to explain the exchange bias training effect in different magnetic system [40-45].

**S13: Out-of-plane training effect**

Figure S13 shows the out-of-plane training of the exchange bias field at 2 K, with the cooling field applied along the corresponding directions. Because of the training effect, a monotonic decrease of the exchange bias field when cycling the magnetic field through consecutive sweeps is seen in both cases. Corresponding coercive field also decreases gradually with increasing *n*. The percentage change of training is calculated using Eqn. S2 and shown in the inset. The drop (about 60%) is more compared to that along the surface plane.

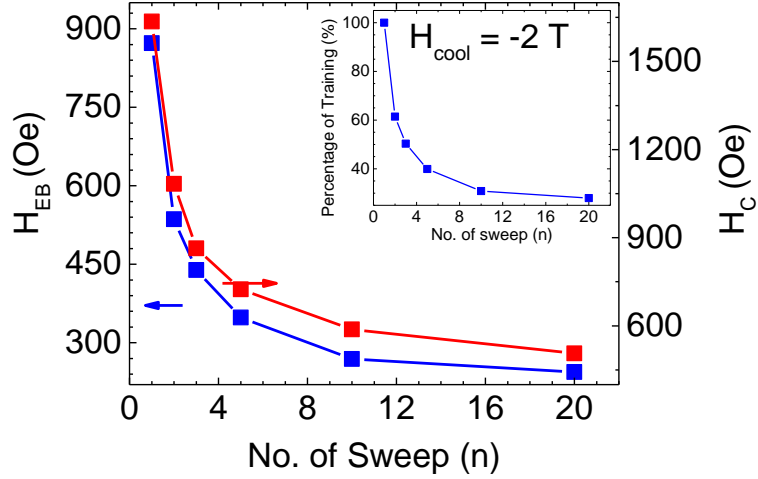

**Figure S13:** Variation of out-of-plane $H_{EB}$ and $H_C$ with number of field cycles ($n$) demonstrating the training effect along the perpendicular orientation. Inset shows the percentage change in the out-of-plane training with $n$.

**S14: Fit to the out-of-plane training effect using Eqn. (4) excluding $n = 1$**

Figure S14 shows fit of out-of-plane training effect data using Eqn. (4) excluding $n = 1$ data point. The obtained value of $H_{EB}(\infty)$ from the fit is about 99.0 Oe that matches very well with the parameters extracted from the fit using Eqn. (6).

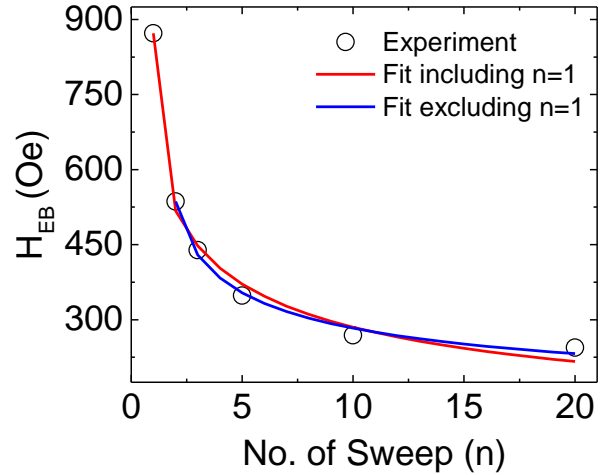

**Figure S14:** Fit (solid line) to the out-of-plane training effect experimental data (solid square) using Eqn. 4 excluding the data point at $n = 1$. The cooling field is –2 T.

**S15: Parameters obtained from the fit to the exchange bias training effect**

(a) **Fitting equations corresponding to different models:**

$$H_{EB}(n) - H_{EB}(\infty) = \frac{k_H}{\sqrt{n}} \qquad (3)$$

$$H_{EB}(n+1) - H_{EB}(n) = \gamma_H [H_{EB}(n) - H_{EB}(\infty)]^3 \qquad (4)$$

$$H_{EB}(n) = H_{EB}(\infty) + A_f \exp\left(-\frac{n}{P_f}\right) + A_i \exp\left(-\frac{n}{P_i}\right) \quad (5)$$

$$H_{EB}(n) = H_{EB}(\infty) + k_H[n + n_0]^{-0.5} \quad (6)$$

**(b) Table 1: Fitting parameters**

| Parameters obtained from | In-plane Training Effect | Out-of-plane Training Effect |
|---|---|---|
| Eqn. (3) | $k_H$ (Oe): 133.2 | $k_H$ (Oe): 618.5 |
|  | $H_{EB}(\infty)$ (Oe): 11.2 | $H_{EB}(\infty)$ (Oe): 86.2 |
|  | $\gamma_H = 1/(2k_H^2)$ ($10^{-7}$ Oe$^{-2}$): 281.9 | $\gamma_H = 1/(2k_H^2)$ ($10^{-7}$ Oe$^{-2}$): 13.1 |
|  | $R^2$: 0.928 | $R^2$: 0.984 |
| Eqn. (4) | $\gamma_H$ ($10^{-7}$ Oe$^{-2}$): 63.9 | $\gamma_H$ ($10^{-7}$ Oe$^{-2}$): 5.8 |
|  | $H_{EB}(\infty)$ (Oe): −22.1 | $H_{EB}(\infty)$ (Oe): 22.8 |
|  | $R^2$: 0.990 | $R^2$: 0.993 |
| Eqn. (4) [Excluding $n = 1$; fitting $\gamma_H$, and $H_{EB}(\infty)$] | $\gamma_H$ ($10^{-7}$ Oe$^{-2}$): 445.6 | $\gamma_H$ ($10^{-7}$ Oe$^{-2}$): 12.8 |
|  | $H_{EB}(\infty)$ (Oe): 23.8 | $H_{EB}(\infty)$ (Oe): 99.0 |
|  | $R^2$: 0.993 | $R^2$: 0.992 |
| Eqn. (5) | $H_{EB}(\infty)$ (Oe): 51.2 | $H_{EB}(\infty)$ (Oe): 243.4 |
|  | $A_f$ (Oe): 71.9 | $A_f$ (Oe): 457.6 |
|  | $P_f$: 3.2 | $P_f$: 3.4 |
|  | $A_i$ (Oe): 1333.2 | $A_i$ (Oe): 2168.1 |
|  | $P_i$: 0.5 | $P_i$: 0.5 |
|  | $P_f/P_i$: 6.4 | $P_f/P_i$: 6.8 |
|  | $R^2$: 0.999 | $R^2$: 0.999 |
| Eqn. (6) | $H_{EB}(\infty)$ (Oe): 23.7 | $H_{EB}(\infty)$ (Oe): 120.3 |
|  | $k_H$ (Oe): 90.9 | $k_H$ (Oe): 492.3 |
|  | $n_0$: −0.9 | $n_0$: −0.6 |
|  | $R^2$: 0.998 | $R^2$: 0.999 |

Fit parameters in Eqn. (5) have been assigned to the interfacial and frozen components assuming that the interfacial rotatable spin component decays faster than the frozen spin component [46,47]. Associated larger pre-factor ($A_i$) of the rotatable component than that of the frozen ($A_f$) could indicate that the training effect is initially dominated by the rotatable spin components. For the in-plane training, the relaxation time ratio, $P_f/P_i \sim 6.48$ indicates that the frozen spins relax about 6 times slower than the rotatable interface spins at 2 K, and the exchange bias is governed mainly by the interface spins. However, a much higher $H_{EB}(\infty) = 51.2$ Oe is obtained from the fit using Eqn. (5), compared to Eqn. (2) and Eqn. (3) (even after excluding $n = 1$ data). A large value of $H_{EB}(\infty)$ and the value $A_i \gg A_f$ associated with $P_i < P_f$ could be due to the nature of exponential fit compared to the power law in Eqn. (3), although it also may point to a different physical picture than the power law dependence.

For the in-plane training, fit parameters obtained using Eqn. (6) are very close to the fit using Eqn. (4) excluding $n = 1$. Excluding $n = 1$ in Eqn. (4), in-plane $H_{EB}(\infty)$ is about 23.8 Oe, and $\gamma_H$ about $445.6 \times 10^{-7}$ Oe$^{-2}$ (corresponding $k_H = 105.9$ Oe). It is mentioned by Sahoo et al. [48] that the discretization method in Binek's model [10] is valid for only small measurement time $\tau$. Since, $2\gamma_H(H_{EB}(n) - H_{EB}(\infty))^2$ becomes smaller for larger $n$, Eqn. (4) predicts correct behavior only for $n > 1$ in our case (see Table 2 and related discussion in supplementary information S16). Eqn. (6) also predicts a positive $H_{EB}(\infty) = 23.7$ Oe which is larger than that predicted by Eqn. (3) but smaller than that by Eqn. (5). Comparing the fits from Eqn. (5) and (6), it is difficult to identify if one of them describes the underlying physical mechanism better than the other ($R^2$ values being the same as listed in Table 1 in S15). More detail theoretical and experimental studies are necessary in this aspect.

### S16: Validity of training effect fit with Eqn. (4)

As described in supplemental section S12, Eqn. (4) could be obtained from Eqn. (6) with the approximation, $2\gamma_H(H_{EB}(n) - H_{EB}(\infty))^2 \ll 1$. Using the fitting parameters obtained from Eqn. (6) we attempt to verify if this approximation is indeed valid for our data. Table 2 lists the measured exchange bias training data points with corresponding value of $2\gamma_H(H_{EB}(n) - H_{EB}(\infty))^2$ listed side by side, where the values of $\gamma_H$ are obtained from $k_H$ using $\gamma_H = \left(\frac{1}{2k_H}\right)^2$, with $k_H$ and $H_{EB}(\infty)$ are noted in Table 1 above. It can be observed that for both in-plane and out-of-plane training, $n = 1$ data points do not satisfy the approximation, $2\gamma_H(H_{EB}(n) - H_{EB}(\infty))^2 \ll 1$. Hence the fit using Eqn. (4) breaks down when the $n = 1$ data point is included.

**Table 2: Validity of Eqn. (4) verified using Eqn. (6)**

|  | $n$ | $H_{EB}(n)$ Oe | $2\gamma(H_{EB}(n) - H_{EB}(\infty))^2$ |
|---|---|---|---|
| In-plane | 1 | 283 | **8.13** |
|  | 2 | 114 | 0.99 |
|  | 3 | 82.5 | 0.42 |
|  | 4 | 71.5 | 0.28 |
|  | 5 | 68.5 | 0.24 |
|  | 8 | 56.0 | 0.13 |
|  | 10 | 54.0 | 0.11 |
|  | 15 | 52.5 | 0.10 |
| Out-of-plane | 1 | 872.5 | **2.33** |
|  | 2 | 536 | 0.71 |
|  | 3 | 439 | 0.42 |
|  | 5 | 348 | 0.21 |
|  | 10 | 269 | 0.09 |
|  | 20 | 244 | 0.06 |

effect. S14: Fit to the out-of-plane training effect using Eq. (4) excluding $n = 1$. S15: Parameters obtained from the fit to the exchange bias training effect. S16: Validity of training effect fit with Eq. (4).